\documentclass [12pt]{article}
\usepackage{graphicx,amssymb,amsfonts,latexsym,amsmath,amsthm,times}
\usepackage{lipsum}
\usepackage{epsfig}
\usepackage{fancyhdr}
\usepackage{titlesec}
\usepackage[english]{babel}
\usepackage[backend=bibtex8,
            bibstyle=ieee,
            sorting=unsrt,
            citestyle=numeric-comp,
            firstinits=true,
            doi=true,
            isbn=false,
            sorting=none,
            url=false]{biblatex}
\usepackage[breaklinks=true]{hyperref}
\hypersetup{colorlinks=true}
\usepackage{breakcites}
\numberwithin{equation}{section}

\newcommand{\pr}{\frac{\partial \rho}{\partial t}}
\newcommand{\rld}[1]{\mathcal{D}_t^{#1}}
\newcommand{\pdx}[1]{\frac{\partial {#1}}{\partial x}}
\newcommand{\pdxs}[1]{\frac{\partial^2 {#1}}{\partial x^2}}

\clearpage
\newtheorem{theorem}{Theorem}
\newtheorem{acknowledgement}[theorem]{Acknowledgement}

\bibliography{ref}
\begin{document}



\vspace*{2cm} \normalsize \centerline{\Large \bf Density-Dependence Subdiffusion in Chemotaxis}

\vspace*{1cm}

\centerline{\bf Akram Al-Sabbagh$^{a,b}$\footnote{Corresponding
author. E-mail: akram.al-sabbagh@postgrad.manchester.ac.uk. This work was done as a part of the author's PhD study at the University of Manchester.}}
\vspace*{0.5cm}

\centerline{$^{a}$School of Mathematics, The university of Manchester, Manchester, Oxford Road, M13 9PL, UK}
\centerline {$^{b}$Dept. of Mathematics and Computer Applications, Al-Nahrrain University, 64055 Baghdad, Iraq}


\vspace*{1cm}
\rule{6in}{1pt}

\centerline{\noindent {\bf ABSTRACT}}
The purpose of this work is to propose a nonlinear non-Markovian model of subdiffusive transport that involves chemotactic substance affecting the cells at all time, not only during the jump. This leads the random waiting time to be dependent on the chemotactic gradient, making the escape rates also dependent on the gradient as well as the nonlinear density dependence. We systematically derive subdiffusive fractional master equation, then we consider the diffusive limit of the fractional master equation. We finally solve the resulting fractional subdiffusive master equation stationery and analyse the role of the chemotactic gradient in the resulting stationary density with a constant and a quadratic chemotactic gradient.

\rule{6in}{1pt}

\vspace*{1cm}
\setcounter{equation}{0}
\section{Introduction}
Cells, bacteria and many other microscopic organisms, move randomly in their environment. That movement plays a key role in many biological phenomena, such as wound healing and embryonic morphogenesis~\cite{fedotov_non-homogeneous_2013}, bacterial aggregation into multicellular masses~\cite{robert_r._kay_changing_2008}, and many other physiological processes~\cite{ridley_cell_2003}.

The random movement results from the response of those living systems to the change of concentration in their environment media. This sort of response is called a \emph{taxis}, which is Greek for ``arrange''~\cite{othmer_aggregation_1997}. Taxis include cell's movement toward or away from from an external situations and it results from changing the cell's movement pattern or direction responding to these external stimulus~\cite{othmer_aggregation_1997}. There are many types of taxis that cell could perform, such as aerotaxis (movement of an organism in response to the presence of molecular oxygen), geotaxis (movement influenced by gravity), haptotaxis (the directional motility or outgrowth of cells) and chemotaxis which we will be studying in this work.

\begin{figure}
	\centering
	\includegraphics[width=0.5\linewidth]{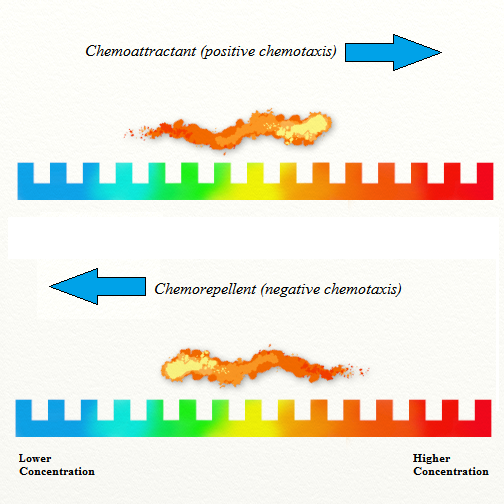}
	\caption[Chemotactic concentration]{The different in organism direction regarding to the chemical signals.}
	\label{fig:chemotaxis}
\end{figure}

Chemotaxis is a guided direct cell movement in a chemical gradient to a higher concentration of beneficial or lower concentration of toxic (favourable environment)~\cite{robert_r._kay_changing_2008}. It also could be referred to be a positive or negative chemotaxis (Figure~\ref{fig:chemotaxis}), depending on the direction of the movement if it is toward or away from the stimulus that affect the chemical gradient~\cite{othmer_aggregation_1997}.

There are two major factors for chemotaxis or any other taxis to be processed. The first is a complex network of signals, such as changing in the global stimulation or chemoattractant concentration~\cite{othmer_aggregation_1997}. Organisms sense those signals, analyse them and then respond. Bacteria, for instance, can sense a wide range of these external signals that is received from the environment~\cite{wadhams_making_2004}. The second component of taxis is the physical response that is appropriate for those signals. The response could be either an alteration in its gene, which will happen as a kind of evolution if these organisms experienced the same signal for long time~\cite{wadhams_making_2004}, or it respond by moving to or away from the source of the signal.

The response in both cases appears in two steps or processes. Cells first detect the signal and then translate or convert those external signal into internal signals which decide what sort of response the cell should make or what direction to move~\cite{othmer_aggregation_1997}. After that, it navigate the direction by examining the gradient of the chemoattractant concentration on its surrounding environment. Cells in general need to direct their movement to specific places to find or avoid some particular stimulus, therefore it is rarely moves unguided~\cite{robert_r._kay_changing_2008}.

The cell movement in chemotactic gradient depends highly on the gradient strength and also on whether the cell itself is polarised or not, it could behave differently in steep gradients in comparison to its behaviour in weak gradients. In strong gradient, for example, cells detect the direction of gradient and and respond directly by active movement, sometime with a little deviation, toward the source of the chemoattractant~\cite{gerisch_chemotactic_1981,zhelev_controlled_2004,swanson_local_1982}. They use receptor occupancy on their surfaces in order to compare and detect where the higher concentration is, this is called directional sensing~\cite{devreotes_eukaryotic_2003}. On the other hand, cells behave differently in a weaker gradient~\cite{robert_r._kay_changing_2008}. In a homogeneous environment for example, cells keep changing their direction randomly about once a second producing an unguided random movement~\cite{wadhams_making_2004}.

Chemotaxis and other cell's random movement have been studied and examined for long time now~\cite{ma_stationary_2012,xue_travelling_2010, painter_continuous_2009, hillen_users_2008, wang_classical_2007, franz_travelling_2013,erban_individual_2004, bournaveas_global_2008, dyson_importance_2014,johnston_mean-field_2012, simpson_exact_2015, perthame_pde_2004}, a huge amount of experimental data has been observed that show important details about chemotaxis pathway, such as the number of molecules of each chemosensory protein within the cell and those proteins' location, and many other information that made it easy to build a mathematical or a computational models which simulate cell movement~\cite{levin_origins_1998}. However, even with the wealth of results and explanations in which these models led to, there are still many unclear aspects~\cite{wadhams_making_2004}. 

In our work, we try to describe the chemotaxis transportation using the continuous time random walk models. CTRW has been widely connected to many physical and biological processes, especially to the field of transportation~\cite{metzler_random_2000,metzler_restaurant_2004}. Many of these transition processes characterise the cell (particle) average mean squared displacement, $\langle x^2(t)\rangle \propto t^\nu$, resulting an anomalous behaviour with subdiffusive if $0<\nu<1$, or superdiffusive for $\nu>1$. 

Subdiffusion is fundamental for many of chemotactic processes, such as bacteria movement~\cite{wadhams_making_2004}, infection response of neutrophils~\cite{peter_j._m._van_haastert_chemotaxis:_2004}, porous media~\cite{drazer_experimental_1999}, cell membranes~\cite{saxton_single-particle_1997,ritchie_detection_2005} and many other subjects.
In the literature, there have been many studies that link chemotaxis with CTRW and concerns random walk that involve run and trap (for more details see~\cite{fedotov_subdiffusion_2011,langlands_fractional_2010,montroll_random_1965,scher_stochastic_1973}). These literature assume that the chemotaxis only affect the particles on the time of jump, and does not affect them as long as they are trapping in their positions.

In order to explain this, let us introduce the classical CTRW model for the evolution of subdiffusing species randomly moving in a one dimensional lattice in a chemoattractant media with gradient of concentration of $\nabla S$. The generalised master equation for such a model is
\begin{equation}\label{mater standard model}
\pr=\lambda(x-l)i(x-l,t)+\mu(x+l)i(x+l,t)-i(x,t),
\end{equation}
where $\rho(x,t)$ is the probability of finding the particle on position $x$ at time $t$, and $i(x,t)$ is and defined by
\begin{equation}\label{standard escape rate}
i(x,t)=\frac{1}{\Gamma(1-\nu)\tau_0^\nu}\rld{1-\nu} \rho(x,t),
\end{equation}
which represents the number of escaping particles from $x$ is the integral escape rate. The integral escape rate $i(x,t)$ in this case does not involve chemotaxis. On the other hand, the jump probabilities to the right $\lambda(x)$ and to the left $\mu(x)$ could depend on chemotaxis gradient as
\begin{equation}\label{standard jump rates}
\lambda(x) =\frac{l}{2} \chi \nabla S,\ \ \ \ \ \ \mu(x) =\frac{l}{2} \chi \nabla S.
\end{equation}
Equation~\eqref{mater standard model} together with~\eqref{standard escape rate} and~\eqref{standard jump rates} leads to the fractional Fokker-Planck equation (FFPE)
\begin{equation}\label{fokker-plank standard}
\pr=D_\nu \left[\pdxs{}-\chi \pdx{}\nabla S \right] \rld{1-\nu}p,
\end{equation}
where $D_\nu$ and $\chi$ refer to the diffusion and chemotaxis coefficients respectively~\cite{metzler_anomalous_1999,metzler_random_2000,barkai_continuous_2000,metzler_deriving_1999}.

In this work, we introduce a nonlinear CTRW model for transport in biological systems with chemotaxis and anomalous subdiffusion, taking into account the effects of chemotactic substance at all of the process time, not only during the jumping. We find the modified fractional diffusion function and evaluate the numerical solution for the stationary density.


\vspace*{0.5cm}
\setcounter{equation}{0}
\section{Nonlinear Non-Markovian Model}

In this section we consider a particle that performs a random walk on a one-dimensional lattice with step size of $l$, assuming that the chemotactic gradient affects this particle all of the time, not only during the jump. That means the random waiting time $T_x$ also depends on the chemotaxis substance $\nabla S$, not only the jumping probabilities as it is the case in model~\eqref{mater standard model}. In order to build our model, we shall use the same mechanism that is used in~\cite{fedotov_subdiffusion_2015}. Assume that the particle makes an instantaneous jump from position $x$ to $x-l$ or $x+l$ after waiting $T_x$. In classic models, when chemotaxis affects the particles during the jump, the jumping mechanism is described as follows: considering the waiting time $T_x^\gamma$ is the minimum of two independent random times $T_x^\lambda$ and $T_x^\mu$, which are the waiting times preceding a jump to the right and left respectively. Then the procedure of jumping is that the particle produces a jump to the right if $T_x^\lambda<T_x^\mu$, and it jumps to the left otherwise.

In this work we consider a more complicated mechanism. Since the chemotaxis gradient affects the particle during staying and jumping, then the assumption now is that there are two types of waiting times, say $T_\gamma$ (which is basically equivalent to $T_x^\gamma$ in the standard model) and $T_c$, which is the additional waiting time that involves chemotaxis. The first waiting time $T_\gamma$ plays the same role in a standard subdiffusive model, where the escape rates $\gamma(x,\tau)=\lambda(x,\tau)+\mu(x,\tau)$ are inversely proportional to the age (residence) time $\tau$. These escape rates could be defined as~\cite{fedotov_non-homogeneous_2013,falconer_nonlinear_2015}
\begin{equation}\label{escape rate anomalous}
\lambda(x,\tau) =\frac{\nu_\lambda(x)}{\tau_0+\tau},\ \ \ \ \  \mu(x,\tau) =\frac{\nu_\mu(x)}{\tau_0+\tau},
\end{equation}
where $\tau_0$ is a parameter with a unit of time, $\nu_\lambda(x)$ and $\nu_\mu(x)$ are the anomalous substance of jumping to the right and left, and the total anomalous coefficient is $\nu(x)=\nu_\lambda(x)+\nu_\mu(x)$. The waiting time $T_\gamma=\min\{T_x^\lambda,T_x^\mu\}$ is drawn from the waiting time PDF $\psi_\gamma(x,\tau)$, that is defined as
\begin{equation}\label{waiting time pdf g}
\psi_\gamma(x,\tau)=\Pr\{\tau<T_\gamma<\tau+d\tau \}=\frac{-\partial \Psi_\gamma(x,\tau)}{\partial \tau},
\end{equation}
where $\Psi_\gamma(x,\tau)$ represents the density with no chemotaxis dependence, which is a product of two survival functions  function $\Psi_\lambda(x,\tau)$ and $\Psi_\mu(x,\tau)$, and defined as
\begin{equation}\label{total survival func g}
\Psi_\gamma(x,\tau)=\Pr\{T_\gamma>\tau \}=\exp\left(\int_0^\tau (\lambda(x,s)+\mu(x,s))ds \right).
\end{equation}
On the other hand, the second mechanism assumes that the escape rate depends on the chemotaxis gradient $\nabla S=S(x-l)-S(x)$. They are defined in~\cite{fedotov_subdiffusion_2015} as
\begin{align}\label{escape rates s}
\lambda_c(x) &=
\begin{cases}
l\chi\nabla S, & \nabla S\geq 0, \\
0, & \nabla S<0,
\end{cases} \notag \\
\mu_c(x) &=
\begin{cases}
0, & \nabla S\geq 0, \\
-l\chi\nabla S, & \nabla S<0.
\end{cases}
\end{align}
The waiting time random variable $T_c$ at the point $x$ is now exponentially distributed, therefore the survival PDF $\Psi_c(x,\tau)$ involves the chemotaxis gradient $\nabla S$, and is defined as
\begin{equation}\label{survival S}
\Psi_c(x,\tau)=e^{-l\chi |\nabla S| \tau},
\end{equation}
where $\chi$ denotes the chemotaxis coefficient of the jump.

At this stage, consider the total residence time at position $x$, denoted by $T_x$, to be the minimum of the two residence times $T_\gamma$ and $T_c$, that is
\begin{equation}\label{residence time total}
T_x=\min\{T_\gamma,T_c \}.
\end{equation}
Hence, the chemotactic gradient prevents the cell from being trapped anomalously at point $x$. In a weak gradient, however, cells seem to move slower and even stay longer at the same point in the absence of chemotaxis~\cite{takeda_role_2007,hoeller_chemotaxis_2007}. This is leading to a huge change in the form of the master equation which describes the model.
In order to derive the generalised master equation, we need first to define the total survival function $\Psi(x,\tau)$ and the total escape rates to the right $\lambda_c(x,\tau)$ and to the left $\mu_c(x,\tau)$ from a point $x$ that involve both chemotaxis and mean field density.
The total survival PDF is a product of the two survival PDFs $\Psi_\gamma(x,\tau)$ and $\Psi_c(x,\tau)$, that is
\begin{align}\label{total survival}
\Psi(x,\tau) &= \Psi_\gamma(x,\tau)\cdot\Psi_c(x,\tau) \notag \\
&= \Psi_\gamma(x,\tau) e^{-l\chi |\nabla S|\tau},
\end{align}
or, by using the definition in~\eqref{total survival func g}
\begin{equation}\label{total survival int(escape)}
\Psi(x,\tau)=\exp\left(-\int_0^\tau (\lambda(x,s)+\mu(x,s))ds-l\chi|\nabla S|\tau \right).
\end{equation}
Recalling the waiting time PDF is $\psi(x,\tau)=-\partial \Psi(x,\tau)/\partial \tau)$, we therefore get
\begin{equation}\label{total waiting time PDF}
\psi(x,\tau)=\psi_\gamma(x,\tau)e^{-l\chi |\nabla S|\tau}+l\chi \nabla S\tau \psi_\gamma(x,\tau)e^{-l\chi |\nabla S|\tau},
\end{equation}
where $\psi_\gamma(x,\tau)$ is given by~\cite{falconer_nonlinear_2015}
\begin{equation}\label{waiting time gamma}
\psi_\gamma(x,\tau)=\psi_\lambda(x,\tau)+\psi_\mu(x,\tau),
\end{equation}
where $\psi_\lambda(x,\tau)=\frac{-\partial \psi_\lambda(x,\tau)}{\partial \tau}\Psi_\mu(x,\tau)$ and $\psi_\mu(x,\tau)=\frac{-\partial \psi_\mu(x,\tau)}{\partial \tau}\Psi_\lambda(x,\tau)$. The general waiting time PDF also depend now on the chemotaxis gradient, involving the exponential tempering factor. Therefore, the tempering in this case is different and more complex.

On the other hand, to define the total nonlinear escape rates to the right and left ($\lambda_c(x,\tau)$ and $\mu_c(x,\tau)$), we take into account the independence of the two mechanisms, so that
\begin{align}\label{escape rates S}
\lambda_c(x,\tau) &=
\begin{cases}
\lambda(x,\tau)+\alpha_\lambda(\rho)+l\chi\nabla S, & \nabla S\geq 0, \\
\lambda(x,\tau)+\alpha_\lambda(\rho), & \nabla S<0,
\end{cases} \notag \\
\mu_c(x,\tau) &=
\begin{cases}
\mu(x,\tau)+\alpha_\mu(\rho), & \nabla S\geq 0, \\
\mu(x,\tau)+\alpha_\mu(\rho)-l\chi\nabla S, & \nabla S<0.
\end{cases}
\end{align}
where $\alpha_\lambda(\rho)$ and $\alpha_\mu(\rho)$ are the additional non-linear escape rates to the right and left, with unspecified dependence 
to $x$ and $t$.
The standard escape rates $\lambda$ and $\mu$ are defined in terms of the both the waiting time and the survival PDFs as
\begin{equation}\label{escape rates gamma (waiting/surv)}
\lambda(x,\tau)= \frac{\psi_\lambda(x,\tau)}{\Psi_\gamma(x,\tau)},\ \ \ \ \ \ \mu(x,\tau)= \frac{\psi_\mu(x,\tau)}{\Psi_\gamma(x,\tau)}.
\end{equation}
The escape rates~\eqref{escape rates S} show that the chemotaxis gradient $\nabla S$ plays a major role in increasing the subdiffusion to the right (if $\nabla S\geq 0$) or to the left (if $\nabla S< 0$). It, however, has no affect on jumping to the left (if $\nabla S\geq 0$) or to the right (if $\nabla S< 0$), in other words, this model shows no symmetry in the part of chemotaxis substance. Also, since the escape rates still depend on the residence time $\tau$, the resulted model involving $\rho(x,t)$ is a non-Markovian. However, we can obtain a Markovian model by suggesting constant escape rates $\lambda$ and $\mu$ and a zero chemotaxis gradient $\nabla S=0$. Also, we can obtain a standard FFPE if we consider a zero chemotaxis coefficient $\chi=0$.

In the next section, we use the technique of structured PDF to evaluate the generalised master equation for the non-Markovian model.


\vspace*{1cm}
\setcounter{equation}{0}
\section{Chemotactic Diffusion and Fractional Equation}

In the literature, numerous attempts can be found to derive a modified fractional diffusion equation that involve chemotaxis substance or any other external forces and reactions~\cite{metzler_random_2000,henry_introduction_2010,fedotov_subdiffusion_2015}. Apart of the work of~\cite{fedotov_subdiffusion_2015}, however, these generalised master equations are not expendable to more general cases of anomalous diffusion, as chemotaxis might have some kind of a physical interaction that links between the time scale of diffusion and the time scale of the effective chemoattractant.

In order to derive the corresponding generalised master equation involving the nonlinear escape rates with chemotaxis dependence defined in~\eqref{escape rates S}, and since the model is non-Markovian, we will use the structured PDF, $\xi(x,t,\tau)$, and add the residence time $\tau$ as an auxiliary variable to the mean field density. This structured density gives the number of particles at state $x$ at time $t$ that have been waiting at $x$ for residence time $\tau$~\cite{falconer_nonlinear_2015,fedotov_non-homogeneous_2013,mendez_reaction-transport_2010}. The PDF obeys the balance equation
\begin{equation}\label{structured balance}
\frac{\partial \xi}{\partial t}+\frac{\partial \xi}{\partial \tau}=-[\lambda_c(x,\tau)+\mu_c(x,\tau)]\xi,
\end{equation}
assuming that the residence time $\tau=0$ at starting time $t=0$, so that the balance equation~\eqref{structured balance} satisfies the initial condition
\begin{equation}\label{initial condition}
\xi(x,0,\tau)=\rho_0(x)\delta(\tau),
\end{equation}
where $\rho_0(x)$ is the initial density. The boundary condition at zero residence time is
\begin{equation}\label{boundary condition}
\xi(x,t,0)=\int_0^t \lambda_c(x-l,\tau)\xi(x-l,t,\tau)d\tau +\int_0^t \mu_c(x+l,\tau)\xi(x+l,t,\tau)d\tau.
\end{equation}

The main target now is to find the general form for the unstructured density
\begin{equation}\label{unstructured density}
\rho(x,t)=\int_0^t \xi(x,t,\tau)d\tau.
\end{equation}
This technique has been used by several authors (see for example~\cite{fedotov_subdiffusion_2011,fedotov_subdiffusive_2012,fedotov_non-homogeneous_2013,mendez_reaction-transport_2010,vlad_systematic_2002,yadav_kinetic_2006}) in order to study the non-Markovian random walk, which is found recently to be a very suitable approach for many nonlinear generalisations~\cite{falconer_nonlinear_2015,fedotov_nonlinear_2013,fedotov_nonlinear_2014,fedotov_subdiffusion_2015,straka_transport_2015}.
First, we solve the balance equation~\eqref{structured balance} using the method of characteristics for $\tau>t$ to get
\begin{equation}\label{characteristics solution}
\xi(x,t,\tau)=\xi(x,t-\tau,0) e^{-\int_0^\tau \gamma(x,s)ds}\  e^{-l\chi|\nabla S|\tau-\int_{t-\tau}^t\alpha(\rho)ds },
\end{equation}
where $\alpha(\rho)=\alpha_\lambda(\rho)+\alpha_\mu(\rho)$. By denoting $j(x,t)=\xi(x,t,0)$ as the integral arrival rate of particles to the point $x$ at exactly time $t$, the structured density in~\eqref{characteristics solution} could be formulated in terms of the arrival rate and the survival PDF~\eqref{total survival} as
\begin{equation}\label{structure (j survival)}
\xi(x,t,\tau)=j(x,t)\Psi_\gamma(x,\tau)e^{-l\chi|\nabla S|\tau}\frac{e^{\Phi(x,t-\tau)}}{e^{\Phi(x,t)}},
\end{equation}
and $\Phi(x,t)=-\int_0^t\alpha(\rho)ds $. It is now convenient to introduce the integral escape rates to the right $i_\lambda(x,t)$ and left $i_\mu(x,t)$ as
\begin{align}\label{integral escape rate}
i_\lambda(x,t) &=\int_0^t \lambda_c(x,\tau) \xi(x,t,\tau)d\tau \notag \\
i_\mu(x,t) &=\int_0^t \mu_c(x,\tau) \xi(x,t,\tau)d\tau.
\end{align}
Considering the singularity of the initial condition~\eqref{initial condition}, and noting that both of the integrals in~\eqref{structure (j survival)} and~\eqref{integral escape rate} involve an upper limit of $\tau=t$, the boundary condition can be rewritten in the form of
\begin{equation}\label{integral arrival rate}
j(x,t) = i_\lambda(x-l,t)+i_\mu(x+l,t) \notag\\
+
\begin{cases}
l\chi\nabla S(x-l)\rho(x-l,t), & \nabla S\geq 0, \\
-l\chi\nabla S(x+l)\rho(x+l,t), & \nabla S<0.
\end{cases}
\end{equation}
Using the characteristic solution in~\eqref{structure (j survival)} together with the initial condition~\eqref{initial condition}, then the integral escape rates can be re-evaluated as
\begin{align}\label{integral escape (char solution) right}
i_\lambda(x,t) &= e^{-\Phi(x,t)}\int_0^t \psi_\lambda(x,\tau) j(x,t-\tau) e^{-l\chi |\nabla S|\tau} e^{\Phi(x,t-\tau)}d\tau \notag\\
&+\psi_\lambda(x,t) \rho_0(x) e^{-l\chi |\nabla S|t-\Phi(x,t)}+\alpha_\lambda(\rho)\rho(x,t),
\end{align}
\begin{align}\label{integral escape (char solution) left}
i_\mu(x,t) &= e^{-\Phi(x,t)}\int_0^t \psi_\mu(x,\tau) j(x,t-\tau) e^{-l\chi |\nabla S|\tau} e^{\Phi(x,t-\tau)}d\tau \notag\\
&+\psi_\mu(x,t) \rho_0(x) e^{-l\chi |\nabla S|t-\Phi(x,t)}+\alpha_\mu(\rho)\rho(x,t).
\end{align}
Also, inserting~\eqref{structure (j survival)} into the expression of the unstructured density~\eqref{unstructured density} and by taking the initial condition~\eqref{initial condition}, we have
\begin{equation}\label{unstructure (char solution)}
\rho(x,t) = e^{-\Phi(x,t)} \int_0^t \Psi_\gamma(x,\tau) j(x,t-\tau) e^{-l\chi|\nabla S|\tau+\Phi(x,t-\tau)}d\tau + \Psi_\gamma(x,t) \rho_0(x) e^{-l\chi|\nabla S|t-\Phi(x,t)}.
\end{equation}
The final step is to find a closed formula for both of the integral escape rates $i_\lambda(x,t)$ and $i_\mu(x,t)$ as well as for the arrival rate $j(x,t)$ in terms of the unstructured density $\rho(x,t)$ in order to find the final formula that obeys the balance equation
\begin{equation}\label{final balance unstructure}
\pr= j(x,t)-i(x,t)-l\chi|\nabla S| \rho(x,t),
\end{equation}
where $i(x,t)$ represents the total escape rate from the point $x$ in which $i(x,t)=i_\lambda(x,t)+i_\mu(x,t)$. Finally, applying the Laplace transform on~\eqref{integral escape (char solution) right}, ~\eqref{integral escape (char solution) left} and~\eqref{unstructure (char solution)} gives the generalised master equation
\begin{align}\label{standard master equation}
\pr &=e^{-\Phi(x,t)} \int_0^t K_\lambda(x-l,t-\tau) e^{-l\chi|\nabla S(x-l)|\tau+\Phi(x,\tau)} \rho(x-l,\tau) d\tau +\alpha_\lambda(\rho(x-l,t))\rho(x-l,t) \notag \\
&+ e^{-\Phi(x,t)} \int_0^t K_\mu(x+l,t-\tau) e^{-l\chi|\nabla S(x+l)|\tau+\Phi(x,\tau)} \rho(x+l,\tau) d\tau +\alpha_\mu(\rho(x+l,t))\rho(x+l,t) \notag \\
&-e^{-\Phi(x,t)} \int_0^t K(x,t-\tau) e^{-l\chi|\nabla S|\tau+\Phi(x,\tau)} \rho(x,\tau) d\tau -\alpha(\rho)\rho(x,t) -l\chi|\nabla S|\rho(x,t) \notag \\
&+
\begin{cases}
l\chi\nabla S(x-l)\rho(x-l,t), & \nabla S\geq 0 \\
-l\chi\nabla S(x+l)\rho(x+l,t), & \nabla S<0
\end{cases}
\end{align}
Here is, $K_\lambda(x,t)$, $K_\mu(x,t)$ and $K(x,t)=K_\lambda(x,t)+K_\mu(x,t)$ are the dependent memory kernels, that defined in the Laplace space
\begin{align}\label{memory kernels}
\hat{K}_\lambda(x,s) &= \frac{\hat{\psi}_\lambda(x,s)}{\hat{\Psi}_\gamma(x,s)} \notag\\
\hat{K}_\mu(x,s) &= \frac{\hat{\psi}_\mu(x,s)}{\hat{\Psi}_\gamma(x,s)}.
\end{align}
We can notice that the master equation~\eqref{standard master equation} contain the exponential factor $e^{-l\chi|\nabla S|\tau}$ that involve the chemotaxis gradient $\nabla S$, beside the exponential factor of the nonlinear rates.


\vspace*{1cm}
\setcounter{equation}{0}
\section{Subdiffusion and the FFPE}

In this section, the fractional master equation, which is a special case of the generalised master equation~\eqref{standard master equation}, will be derived. As mentioned previously, the escape rates $\lambda_c$ and $\mu_c$ are inversely proportional to the residence time; this means it is less likely the cell moves for longer it stays at point $x$. Recalling the definition of the escape rates in~\eqref{escape rate anomalous}, the escape rates and the anomalous exponent are space dependent, and hence the non-Markovian model is also non-homogeneous. The use of~\eqref{escape rate anomalous} in the definition of the survival density~\eqref{total survival func g} leads to the power law dependence survival PDF
\begin{equation}\label{survival anomalous}
\Psi_\gamma(x,\tau)=\left[\frac{\tau_0}{\tau_0+\tau} \right]^{\nu(x)},
\end{equation}
where $\nu(x)$ is the total anomalous exponent that is $\nu(x)=\nu_\lambda(x)+\nu_\mu(x)$. Then the waiting time PDF has a Pareto form
\begin{equation}\label{waiting time pdf anomalous}
\psi_\gamma(x,\tau)=\left[\frac{\nu(x) \tau_0^{\nu(x)}}{(\tau_0+\tau)^{1+\nu(x)}}\right].
\end{equation}

At this point, let us introduce the jump probabilities to the right and left that depend on the anomalous exponent and are independent of the residence time $\tau$~\cite{klages_anomalous_2008,mendez_reaction-transport_2010},
\begin{align}\label{jump prob}
p_\lambda(x) &=\frac{\lambda(x,\tau)}{\gamma(x,\tau)} =\frac{\nu_\lambda(x)}{\nu(x)}, \notag \\
p_\mu(x) &=\frac{\mu(x,\tau)}{\gamma(x,\tau)} =\frac{\nu_\mu(x)}{\nu(x)}.
\end{align}
Using the Tauberian theorem, the memory kernels~\eqref{memory kernels}, can be asymptomatically approximated in the Laplace space as $s\rightarrow 0$ by
\begin{equation}\label{memory kernels anomalous}
\hat{K}_\lambda(x,s) =\frac{p_\lambda(x) s^{1-\nu(x)}}{g(x)},\ \ \ \ \ \hat{K}_\mu(x,s) =\frac{p_\mu(x) s^{1-\nu(x)}}{g(x)},
\end{equation}
where $g(x)=\Gamma(1-\nu(x))\tau_0^{\nu(x)}$. Since $p_\lambda(x)+p_\mu(x)=1$ then the combined memory kernel is
\begin{equation}\label{total memory kernal anomalous}
\hat{K}(x,s)=\frac{s^{1-\nu(x)}}{g(x)}.
\end{equation}
Hence the integral escape rates~\eqref{integral escape (char solution) right} and~\eqref{integral escape (char solution) left} can be rewritten in terms of the memory kernels with anomalous exponent dependence~\eqref{memory kernels anomalous} and~\eqref{total memory kernal anomalous} in the form
\begin{align}\label{escape rates anomalous}
i_\lambda(x,t) &=a(x) e^{-l\chi|\nabla S|t-\Phi(x,t)} \rld{1-\nu(x)}\left[ \rho(x,t) e^{l\chi|\nabla S|t+\Phi(x,t)} \right]+\alpha(\rho)_\lambda\rho(x,t),\notag \\
i_\mu(x,t) &=b(x) e^{-l\chi|\nabla S|t-\Phi(x,t)} \rld{1-\nu(x)}\left[ \rho(x,t) e^{l\chi|\nabla S|t+\Phi(x,t)} \right]+\alpha_\mu(\rho)\rho(x,t),
\end{align}
where $a(x)=\frac{p_\lambda(x)}{g(x)}$ and $b(x)=\frac{p_\mu(x)}{g(x)}$ are the anomalous rate coefficients, and $\rld{1-\nu(x)}$ represents the Riemann-Liouville fractional derivative.

The aim is to evaluate the fractional chemotaxis master equation for cell movement in a diffusive case. Therefore, after inserting the diffusive escape rates~\eqref{escape rates anomalous} into the generalised master equation~\eqref{final balance unstructure}, we expand the resulting formula using Taylor expansion to second order in step size $l$ to get
\begin{align}\label{fractional diffusion master general}
\pr &=-l \pdx{}\left[ i_\lambda(x)-i_\mu(x)+l\chi \nabla S \rho(x,t) \right]+\frac{l^2}{2} \pdxs{}i(x,t).
\end{align}
For specific jump probabilities $p_\lambda(x)$ and $p_\mu(x)$ (for more details see~\cite{fedotov_non-homogeneous_2013,falconer_nonlinear_2015}), the difference in $i_\lambda(x,t)$ and $i_\mu(x,t)$ can be approximated by the use of
\begin{equation}
p_\lambda(x)-p_\mu(x)\approx -l \beta \nabla S,\ \ \ \ \ p_{\alpha,\lambda}(x,t)-p_{\alpha,\lambda}(x,t)\approx -l\kappa \pdx{\rho},
\end{equation}
and therefore the fractional diffusion equation can take the form
\begin{align}\label{fractional diffusion master}
\pr &= -l^2 \pdx{}\left[\frac{\beta \nabla S}{g(x)} e^{-l\chi|\nabla S|t-\Phi(x,t)} \rld{1-\nu(x)} (\rho(x,t) e^{l\chi|\nabla S|t+\Phi(x,t)})+\chi \rho(x,t)\nabla S+\kappa \pdx{\rho} \right] \notag \\
&+\frac{l^2}{2} \pdxs{}i(x,t).
\end{align}
In particular, for equal  jump probabilities $p_\lambda(x)=p_\mu(x)=1/2$ we get equal anomalous rate coefficients $a(x)=b(x)$ and equal nonlinear rates $\alpha_\lambda(\rho)=\alpha_\lambda(\rho)$ and therefore an equal escape rates $i_\lambda(x,t)=i_\mu(x,t)$. Hence, the fractional master equation can be written in form
\begin{align}\label{fractional diffusion master ir=il}
\pr &=-l^2 \chi \pdx{}\Big[\nabla S \rho(x,t) \Big] +\frac{l^2}{2}\pdxs{}i(x,t).
\end{align}
The fractional master equation~\eqref{fractional diffusion master ir=il} can be described as a transition adapter that can switch between transition modes regarding to the waiting time of jumps. It obtains an intermediate subdiffusive mode when $\tau_0/T_1\ll 1$ and $l\chi|\nabla S|T_1\ll 1$, whereas a normal diffusion is observed if $l\chi|\nabla S|T_1\gg 1$. We also notice that equation~\eqref{fractional diffusion master ir=il} has no drift, instead it involves the chemotaxis gradient in both terms of the right-hand side. Chemotaxis does not only determine the advection, as in the case of the standard model~\eqref{fokker-plank standard}, but it also involves a tempering parameter through the exponential factor $e^{l\chi|\nabla S|t}$.

The next step is to find a stationary density $\rho_{st}(x)$ that can approximate the fractional diffusion equation~\eqref{fractional diffusion master general} and for the special case~\eqref{fractional diffusion master}. The solution for equation~\eqref{fractional diffusion master ir=il} has already been found in~\cite{fedotov_subdiffusion_2015}, but only for the linear case.


\vspace*{1cm}
\setcounter{equation}{0}
\section{Stationary FFPE}

The aim of this section is to evaluate the stationary solution of the generalised fractional master equation~\eqref{fractional diffusion master general} and~\eqref{fractional diffusion master}. In order to derive their formulas we apply the Laplace transform to the integral escape rates~\eqref{integral escape (char solution) right} and~\eqref{integral escape (char solution) left}, by the use of the shift theorem, we get
\begin{align}\label{integral escape rate laplace}
\hat{i}_\lambda(x,s) &=a(x) [s+\alpha(\hat{\rho})+l\chi|\nabla S|]^{1-\nu(x)} \hat{\rho}(x,s)+\alpha_\lambda(\hat{\rho})\hat{\rho}(x,s), \notag \\
\hat{i}_\mu(x,s) &=b(x) [s+\alpha(\hat{\rho})+l\chi|\nabla S|]^{1-\nu(x)} \hat{\rho}(x,s) +\alpha_\mu(\hat{\rho})\hat{\rho}(x,s).
\end{align}
The stationary density is defined in the limit of $s\rightarrow 0$, corresponding to $t\rightarrow \infty$ as
\begin{equation}\label{stationary definition}
\rho_{st}(x)=\lim_{s\rightarrow 0} s \hat{\rho}(x,s),
\end{equation}
and therefore, the stationary escape rates are
\begin{align}\label{stationary escape rate definition}
i_{\lambda,st}(x) &=\lim_{s\rightarrow 0} s \hat{i}_\lambda(x,s), \notag \\
i_{\mu,st}(x) &=\lim_{s\rightarrow 0} s \hat{i}_\mu(x,s),
\end{align}
hence, the stationary escape rates will take the Markovian form
\begin{align}\label{stationary integral escape rate}
i_{\lambda,st}(x) &=\Bigg[ a(x) \Big[\alpha(\rho_{st})+\chi|\nabla S|\Big]^{1-\nu(x)} +\alpha_\lambda(\rho_{st})\Bigg] \rho_{st}(x), \notag \\
i_{\mu,st}(x) &=\Bigg[ a(x) \Big[\alpha(\rho_{st})+\chi|\nabla S|\Big]^{1-\nu(x)} +\alpha_\mu(\rho_{st})\Bigg] \rho_{st}(x),
\end{align}
and the total stationary escape rate, since $a(x)+b(x)=1/g(x)$, is
\begin{equation}\label{total stationary escape rate}
i_{st}(x) =\left[ \frac{\Big[\alpha(\rho_{st})+\chi|\nabla S|\Big]^{1-\nu(x)}}{g(x)}  +\alpha(\rho_{st})\right] \rho_{st}(x).
\end{equation}
For simplicity, we can write the integral escape rates as
\begin{equation}\label{integral escape rates (nu)}
i_{\lambda,st}(x) =\lambda_\nu(x) \rho_{st}(x),\ \ \ \ \ 
i_{\mu,st}(x) =\mu_\nu(x) \rho_{st}(x),
\end{equation}
where
\begin{align}
\lambda_\nu(x) &=a(x) \Big[\alpha(\rho_{st})+l \chi |\nabla S| \Big]^{1-\nu(x)}+\alpha_\lambda(\rho_{st}), \notag \\
\mu_\nu(x) &=b(x) \Big[\alpha(\rho_{st})+l \chi |\nabla S| \Big]^{1-\nu(x)}+\alpha_\mu(\rho_{st})
\end{align}
Expanding the formula~\eqref{fractional diffusion master general} with respect to $l$ for the second order, with the time derivative of the density approaches zero $\pr\rightarrow 0$, and by the use of~\eqref{integral escape rates (nu)} leads to the final form of the stationary master equation can take the form
\begin{equation}\label{final stationary density}
\pdxs{} D_\nu(x,\rho_{st})\rho_{st}(x)= \pdx{}v_\nu(x,\rho_{st}) \rho_{st}(x),
\end{equation}
where $D_\nu(x,\rho_{st})$ is the diffusion coefficient, and $v_\nu(x,\rho_{st})$ is the velocity, and that are defined as
\begin{align}
D_\nu(x,\rho_{st}) &= \frac{l^2}{2} \frac{\Big[\alpha(\rho_{st})+l \chi |\nabla S| \Big]^{1-\nu(x)}}{g(x)}+\alpha(\rho_{st}) \label{diffusion co}\\
v_\nu(x,\rho_{st}) &= \frac{l \beta \nabla S}{g(x)} \Big[\alpha(\rho_{st})+l \chi |\nabla S| \Big]^{1-\nu(x)} +l \kappa \pdx{\rho_{st}} \alpha(\rho)+l \chi \nabla S.\label{velocity co}
\end{align}

The stationary solution~\eqref{final stationary density} is not Blotzmann distributed as the case of the standard model, the reason is that the diffusion coefficient depends now on the chemotaxis gradient as well as the anomalous exponent, therefore (comparing to the Boltzmann case) cells now spread further and faster.

Fedotov and Korabel have found that for large $x$ and $\nabla S=-kx$, the stationary density $\rho_{st}(x)$~\eqref{stationary density general ir=il} is approximated by
\begin{equation}\label{stationary density final ir=il}
\rho_{st}(x)\sim e^{-A |x|^{1-\nu(x)}},
\end{equation}
with $A>1$ is a constant (for more details see~\cite{fedotov_subdiffusion_2015}).

\begin{figure}[htp]
	\begin{center}
		\includegraphics[width=3in]{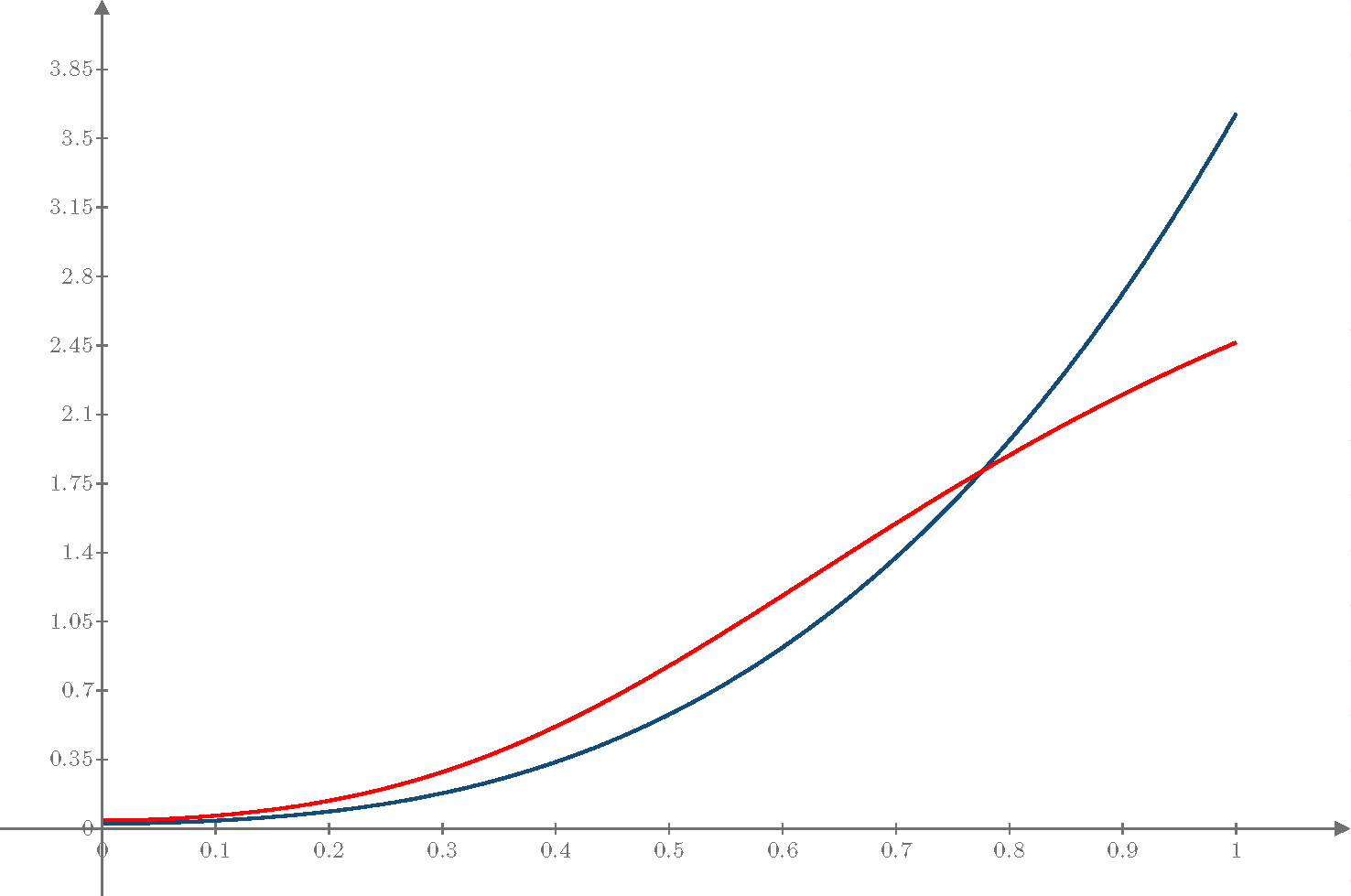}
		\caption{Stationary density of particles in constant chemotaxis gradient $\nabla S=0.5$ (blue line) and $\nabla S= 0.001$ red line, calculated with anomalous substance $\nu(x)=0.9 e^{-k x}$, with parameters $k=2.19$, $\chi=0.1$ and a equal jump probabilities $p_\lambda(x)=p_\mu(x)=0.5$ and an equal nonlinear escape rates $\alpha_\lambda(\rho)=\alpha_\mu(\rho)=(1+0.1\rho)\times 10^-2 $.}
		\label{linear chemotaxis}
	\end{center}
\end{figure}

\begin{figure}[htp]
	\begin{center}
		\includegraphics[width=3in]{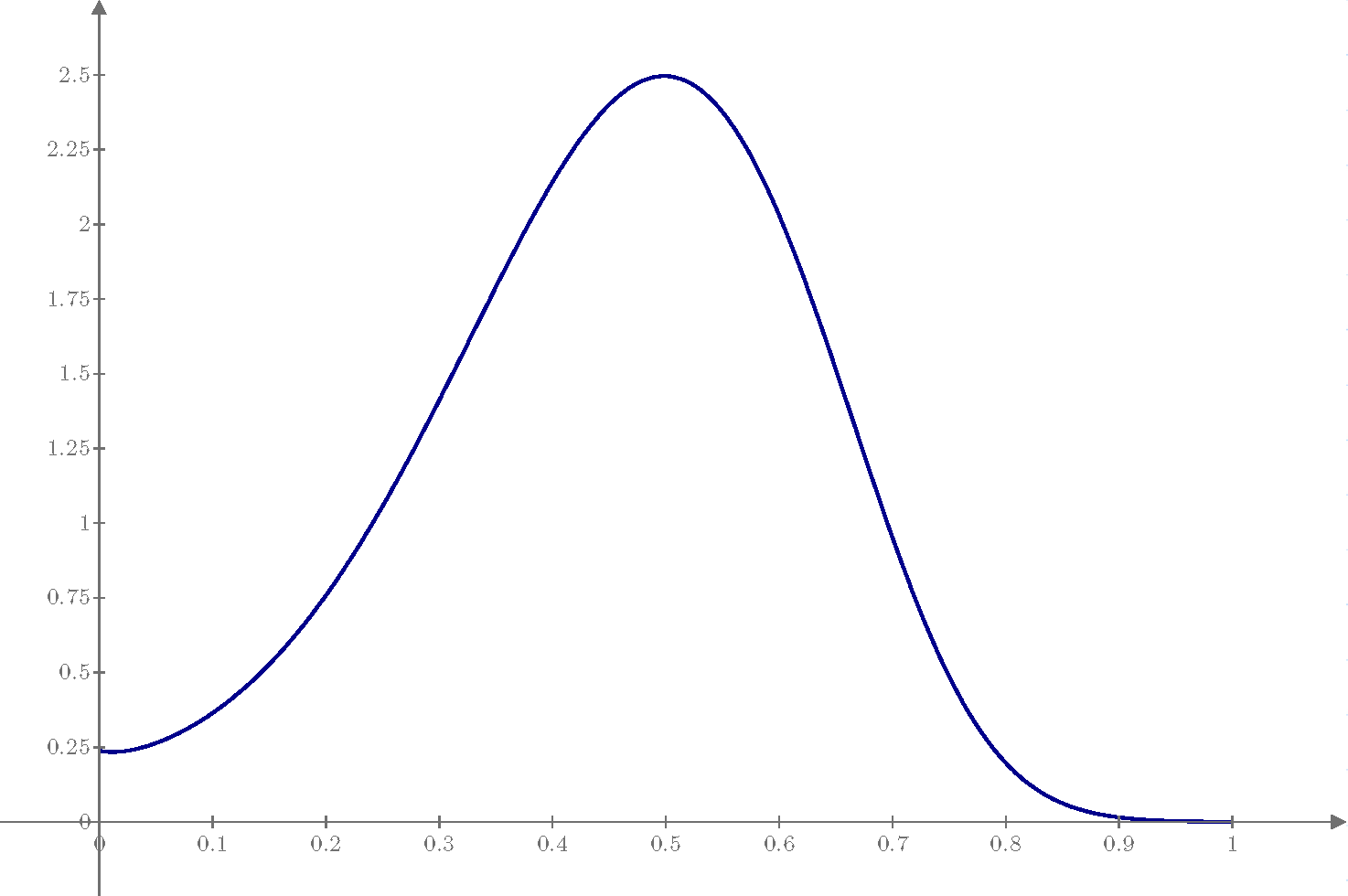}
		\caption{Stationary density of particles in quadratic chemotaxis gradient $\nabla S=kx^2$, calculated with anomalous substance $\nu(x)=0.9 e^{-kx}$ where $k=2.19$, $\chi=0.1$, with an equal probability of jump $p_\lambda(x)=p_\mu(x)=0.5$ and an equal nonlinear escape rates $\alpha_\lambda(\rho)=\alpha_\mu(\rho)=(1+0.1\rho)\times 10^-2 $.}
		\label{nonlinear chemotaxis}
	\end{center}
\end{figure}

Our attempt is to find an approximated solution for the stationary density $\rho_{st}$ by solving \eqref{final stationary density} numerically. We managed to solve it with some assumptions that make it easier to deal with. Otherwise, it is clear that it is very hard to simulate because of the dependence on the local density and the chemotactic gradient in the same time. Therefore, we assume an equal jump probability to the right and left $p_\lambda=p_\mu$, and also an equal nonlinear escape rates for the process $\alpha_\lambda=\alpha_\mu$. therefore, the velocity coefficient~\eqref{velocity co} will be reduced to the form of
\[v_\nu(x,\rho_{st})=l \chi \nabla S \].
Two cases are considered in evaluating those results. The firs is for a constant chemotactic gradient $\nabla S=0.5 \text{ and } 0.001$, Figure~\eqref{linear chemotaxis} shows that for steep gradient $\nabla S=0.5$, particles spread farther and faster, whereas in a weaker gradient $\nabla S=0.001$, particles move slower and trap longer. On the other hand, for non-constant decreasing chemotactic gradient $\nabla S=-k x^2$, particles seem to aggregate where the gradient level is in the average. The result of the second case is presented in Figure~\eqref{nonlinear chemotaxis}. Then, one can notice that cells avoid moving to the region where the chemoattractant level is very high or very low, otherwise, cells aggregate in the average concentration area.

\vspace*{0.5cm}

\section{Conclusions}
In this work we studied a non-Markovian random walk model in which chemotaxis affects cells (particles) at all time, during  jumping and trapping. This means that the waiting time~\eqref{residence time total} variable and also the escape rates~\eqref{escape rates S} depends on the chemotactic gradient $\nabla S$. We presented the generalised master equation~\eqref{standard master equation} which include the exponential factor involving the chemotactic gradient as well as the nonlinear rates. The chemotaxis prevent cells to be trapped anomalously at their position.
In the diffusive limit when $l\rightarrow 0$, we had a new type of fractional diffusive equation~\eqref{fractional diffusion master general}. For an equal jump probabilities, one can notice that the fractional equation has no drift, otherwise chemotaxis appears in both terms instead. It is different from the standard fractional diffusion by involving the chemotaxis gradient not only in the drift term, but it also appears in the diffusion term. We then obtained the stationary solution and presented the results with two cases. The first case is where the chemotaxis gradient is constant, that led to an anomalous in a sense of that the diffusion coefficient depends on the gradient and also on the anomalous substance. results showed that cells in weak chemotactic gradient tend to be moving slowly and trap for longer, in contrast of strong gradient where cells spread widely. The second case involved a quadratic chemotaxis gradient. In this case, cells avoid moving to the region where the chemoattractant level is very high or very low, otherwise, cells aggregate in the average concentration area. Results were shown in figures~\eqref{linear chemotaxis} and~\eqref{nonlinear chemotaxis}. This Model can also be applied on transport systems with non-standard diffusion, where memory kernels causes a dependence between reactions and transport, like the case of propagating front~\cite{yadav_propagating_2007}.

\begin{acknowledgement}
 The author is very grateful to Prof. Sergei Fedotov for his great help and support.
\end{acknowledgement}


\printbibliography

\end{document}